\documentclass[twocolumn,aps,pra,superscriptaddress,floatfix,showpacs,notitlepage]{revtex4-1}
\usepackage[latin9]{inputenc}
\usepackage{color}
\usepackage{textcomp}
\usepackage{amsmath}
\usepackage{amssymb}
\usepackage{graphicx}
\usepackage{esint,hyperref}
\usepackage{physics}
\usepackage[inline]{enumitem}

\begin{document}
\title{Machine learning assisted readout of trapped-ion qubits} 
\author{Alireza Seif}
\affiliation{Joint Quantum Institute and Department of Physics, University of Maryland, College Park, MD 20742}
\author{Kevin A. Landsman}
\affiliation{Joint Quantum Institute and Department of Physics, University of Maryland, College Park, MD 20742}
\affiliation{Joint Center for Quantum Information and Computer Science, University of Maryland, College Park, MD 20742}
\author{Norbert M. Linke}
\affiliation{Joint Quantum Institute and Department of Physics, University of Maryland, College Park, MD 20742}
\affiliation{Joint Center for Quantum Information and Computer Science, University of Maryland, College Park, MD 20742}
\author{Caroline Figgatt}
\affiliation{Joint Quantum Institute and Department of Physics, University of Maryland, College Park, MD 20742}
\affiliation{Joint Center for Quantum Information and Computer Science, University of Maryland, College Park, MD 20742}
\author{C. Monroe}
\affiliation{Joint Quantum Institute and Department of Physics, University of Maryland, College Park, MD 20742}
\affiliation{Joint Center for Quantum Information and Computer Science, University of Maryland, College Park, MD 20742}
\affiliation{IonQ, Inc., College Park, MD 20740 USA} %This one's for Chris
\author{Mohammad Hafezi}
\affiliation{Joint Quantum Institute and Department of Physics, University of Maryland, College Park, MD 20742}
\affiliation{Department of Electrical and Computer Engineering and Institute for Research in Electronics and Applied Physics, University of Maryland, College Park, MD 20742}
\date{\today}
\begin{abstract}
We reduce measurement errors in a quantum computer using machine learning techniques. We exploit a simple yet versatile neural network to classify multi-qubit quantum states, which is trained using experimental data. This flexible approach allows the incorporation of any number of features of the data with minimal modifications to the underlying network architecture. We experimentally illustrate this approach in the readout of trapped-ion qubits using additional spatial and temporal features in the data. Using this neural network classifier, we efficiently treat qubit readout crosstalk, resulting in a 30\% improvement in detection error over the conventional threshold method. Our approach does not depend on the specific details of the system and can be readily generalized to other quantum computing platforms.

\end{abstract}
\maketitle

Quantum computing tasks  involve quantum state preparation, time evolution, and measurement, accompanied by errors in all the three stages. To detect and correct errors during the evolution, quantum error correcting codes are used\cite{Shor95,Steane1996multiple,gottesman2009introduction}. These codes rely on redundant encoding of quantum information, which makes it possible to measure syndromes and fix errors. Measurement errors not only affect the outcome of the computation, but they also limit the task of error correction. Consequently, in addition to high-fidelity operations, high quality multi-qubit readout is essential for realizing a fault tolerant quantum computer. 
 \begin{figure}[t]
	\includegraphics[width=1\linewidth]{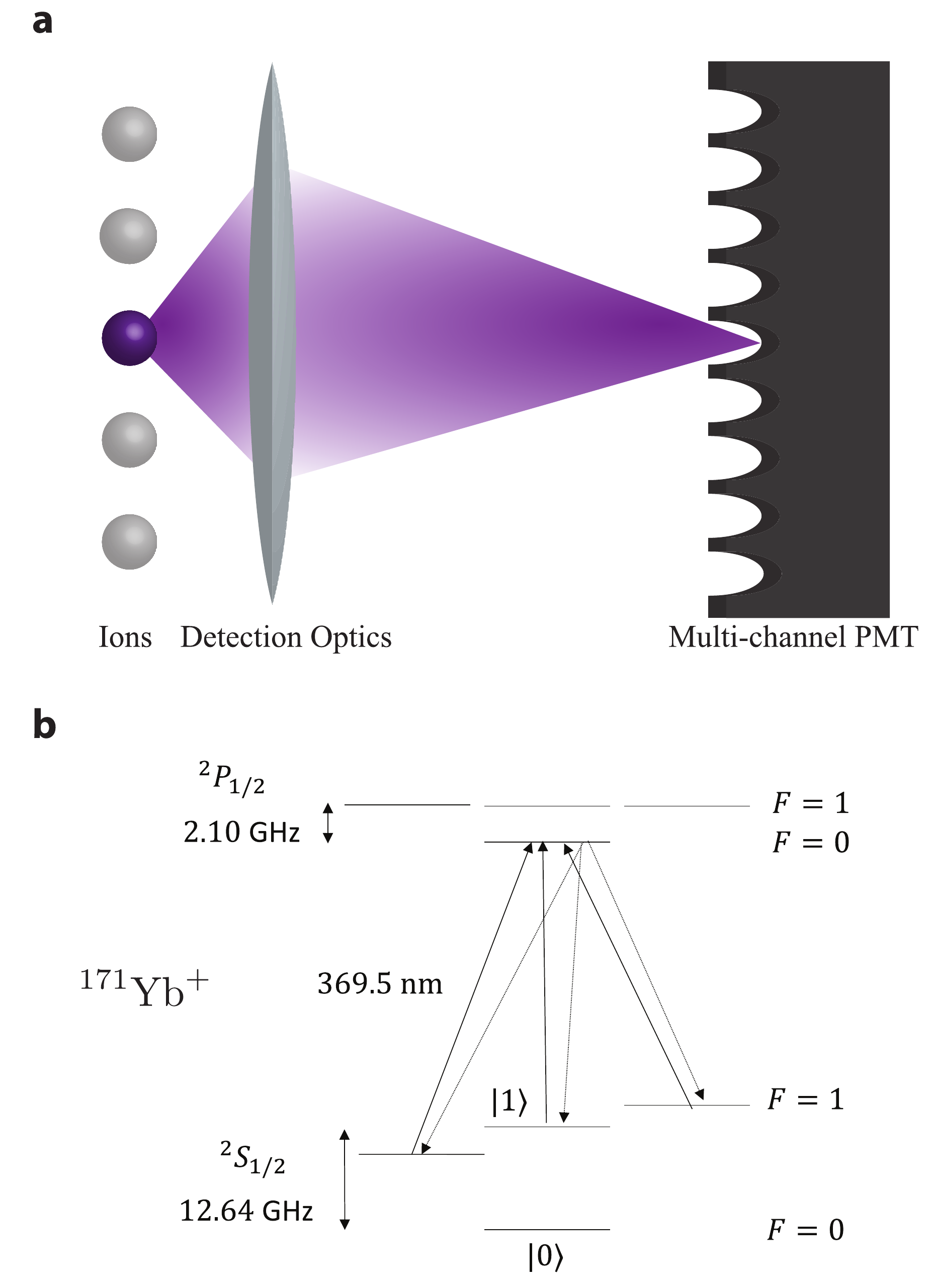}
	\caption{The setup and readout scheme for the trapped-ion quantum computer. (a) Schematic of our experimental setup. A single ion fluoresces inside an ion trap and its radiated photons are collected by a 0.37 NA lens. This fluorescence is then imaged onto a single channel of a multi-channel photo-multiplier tube detector. (b) Energy levels in the $^{171}\rm{Yb}^+$ atomic system used for fluorescence detection. If the qubit state is $\ket{0}$, the applied detection laser is off resonance and nearly no photons are scattered. If the state is $\ket{1}$ the transition is on resonance and the ion fluoresces strongly. }
	\label{fig:yblevels}
\end{figure}

The quantum measurement process always involves the interaction with an external classical system. For example, collecting fluorescence from a trapped ion in a cycling transition can determine the state of the qubit \cite{myerson2008high}. In superconducting qubits, a probe signal is injected to the system through a resonator, and the phase of the output signal is used to infer the state of the qubit \cite{jeffrey2014fast}. Spontaneous decay and excitation during the external probe can be major sources of qubit measurement errors
\cite{olmschenk2007manipulation}.
When scaling up, the measurement signal from a qubit can be  altered by the state of other qubits through crosstalk. To address this issue, one can assume an error model and infer the correct qubit states by using statistical properties of the measured data.\cite{burrell2010high}. 

Machine learning (ML) \cite{jordan2015machine} techniques have recently become popular tools for exploring physical phenomena. For example, artificial neural networks \cite{nielsen2015neural} are now a powerful method for simulating the dynamics of many-body quantum systems \cite{carleo2017solving,bosehubbard2017}. These neural networks can efficiently represent a wide class of highly correlated states \cite{gao2017efficient,deng2017quantum,deng2016exact}, and can facilitate quantum state tomography \cite{Torlai2017,Torlai2018}. They are also used to detect errors and decode quantum error correcting codes \cite{baireuther2017machine,krastanov2017deep,torlai2017neural}, and to classify phases of matter \cite{van2017learning,van2017learning2,carrasquilla2017machine}. In addition to neural networks, other ML methods, such as principal component analysis and clustering, have been used for various tasks from classifying phases of matter \cite{wetzel2017unsupervised} to discriminating measurement trajectories for improved single-qubit readout\cite{magesan2015machine}.

In this work, we exploit the versatility of ML techniques to increase the fidelity of multi-qubit measurements. While the problem of crosstalk can be partially addressed by careful statistical analysis of the data, it requires certain assumptions about the error model, which makes the integration of additional spatial and temporal features difficult. In our approach, the machine is ``trained" to infer the states from the measurement results without prior knowledge of the error model. This ML method can therefore be readily generalized to other quantum computing platforms.

We study the detection accuracy of a chain of $^{171}\text{Yb}^+$ ions confined in an rf Paul trap \cite{debnath2016demonstration}. The qubit is defined by the hyperfine-split ground states of the $^2\text{S}_{1/2}$ manifold: $\ket{0} = \ket{F=0,m_F=0}$ and $\ket{1} = \ket{F=1,m_F=0}$ (see Fig.~\ref{fig:yblevels}). Furthermore, we can take advantage of the $^2\text{P}_{1/2}$ level to accomplish both state preparation and measurement (SPAM) with high fidelity. 

\begin{figure}[t]
	\includegraphics[width=1\linewidth]{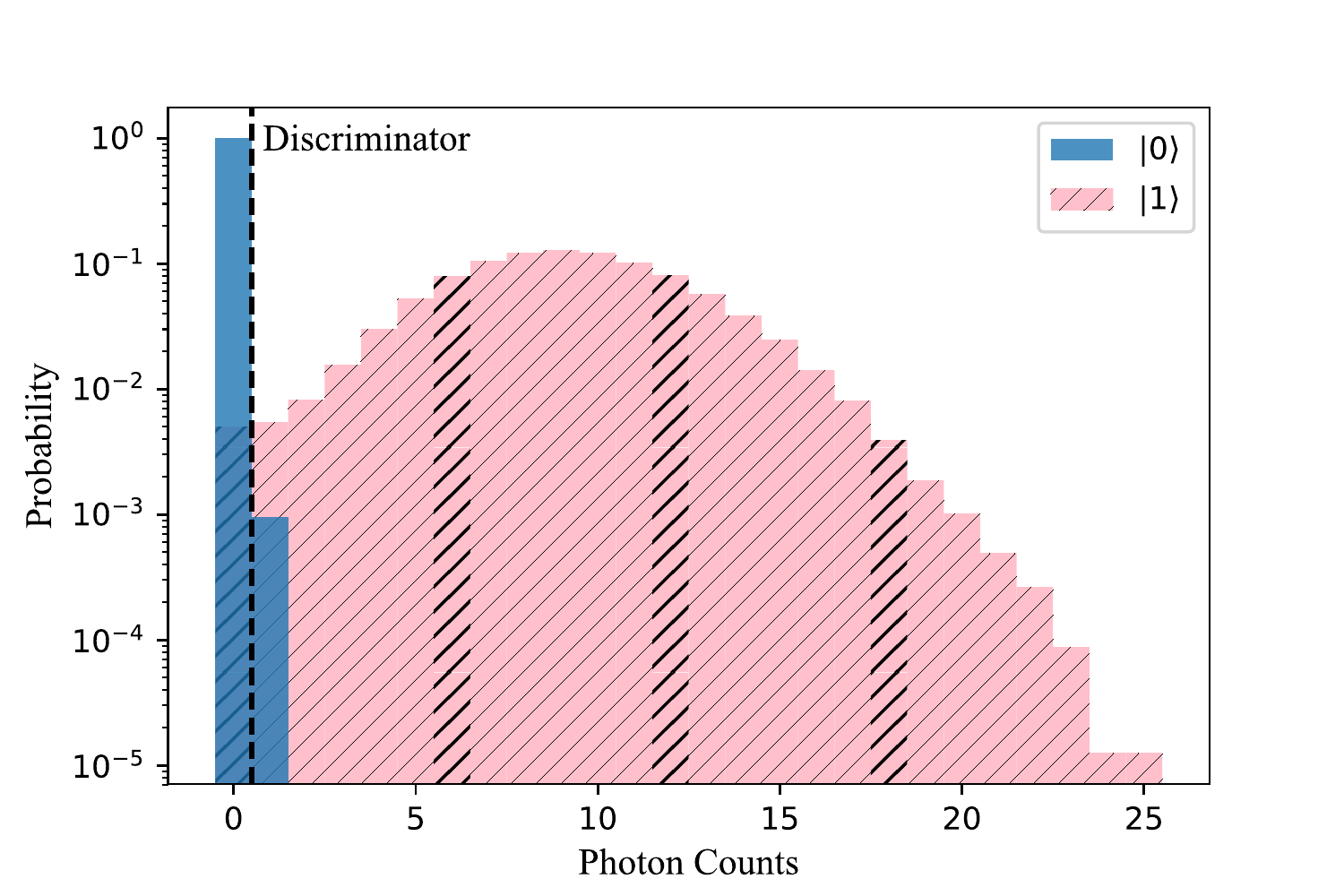}
	\caption{The histogram of observed photons for an integration time of 150 $\mu$s for state detection. The photon counts follow a Poisson distribution, in which the state $\ket{0}$ (solid blue) gives a mean close to zero, while the state $\ket{1}$ (shaded pink) results in nine photons on average.}
	\label{fig:histogram}
\end{figure}

Qubit initialization is achieved by optical pumping via the $\ket{^2 P_{1/2},F=1}$ manifold. The qubit readout, on the other hand, is performed  by state-dependent fluorescence detection \cite{olmschenk2007manipulation} (see Fig.~\ref{fig:histogram}). Specifically, we  apply a laser beam resonant with the $\ket{^2 S_{1/2},F=1} \to \ket{^2 P_{1/2},F=0}$ cycling transition, and collect ion fluorescence. While the beam is on, a qubit in $\ket{1}$ will scatter photons. In contrast, a qubit in $\ket{0}$ remains dark since the light is 14.7 GHz detuned from the nearest transition with a natural linewidth of about 20 MHz. The ion fluorescence is collected by a 0.37 NA lens and each ion in the chain is imaged onto a separate channel of a 32-channel photo-multiplier tube (PMT) \cite{debnath2016demonstration}. 

The histogram of the photon counts in some integration time follows a near-Poissonian distribution, centered around 0 for state $\ket{0}$  (the ``dark" state) and 9 counts for state $\ket{1}$ (the ``bright" state)  following a 150 $\mu$s integration time.  The deviations from Poissonian statistics  indicate the error mechanisms in this readout scheme. The dark state histogram includes a small contribution at higher counts due to off-resonant dark-to-bright pumping during the detection step \cite{acton2006near}. More importantly, the bright state histogram has a non-Poissonian tail towards lower photon counts due to off-resonant excitation to the $\ket{^2 P_{1/2}, F=1}$ manifold, detuned by 2.1 GHz, from which decay to $\ket{0}$ is possible\cite{acton2006near}. By choosing an optimal collection time, 150 $\mu$s in our system, the overlap between the photon distributions corresponding to $\ket{0}$, and $\ket{1}$ can be minimized. Thus, by discriminating the two distributions one can deduce the state of the qubit. One of the commonly used techniques to distinguish between these distributions is a simple threshold discriminator, where instances with observed photon counts greater than the threshold are taken to be $\ket{1}$, and those below to be $\ket{0}$ (see Fig.~\ref{fig:histogram}). This method works very well in the single qubit case and results in a detection fidelity, that is $\mathcal{F} = p(\text{measured x}|\text{prepared x})$, of $99.4\%$ for $\ket{1}$, and $99.6\%$ for $\ket{0}$, which gives an average detection fidelity of $99.5\%$ in our setup. This error can be reduced by increasing the collection angle of the objective and reducing the detection time. A readout fidelity of $\sim99.99\%$ has been demonstrated using this method \cite{crain2016int}.

Similar dark-to-bright or bright-to-dark pumping errors exist in other readout schemes, e.g., when a separate state with a finite lifetime is used as the dark state, known as a shelf state \cite{burrell2010high}. Smaller  error contributions include laser light scattering off the ion trap and  into the PMT as well as PMT dark counts , which account for 20 counts per second and 2 counts per second respectively. Both of these errors contribute only one false count for an average of 300 experiments and are therefore well discriminated using a thresholding method.

When detecting the state of more than one qubit, a bright ion can cause events on other ion detector channels. This crosstalk between the PMT channels modifies the distribution of observed photons, and the average detection fidelity decreases. One can choose a different threshold for each ion based on the state of its neighbors to partially mitigate these errors. In addition, using  maximum likelihood methods, one can calculate the probability that an observed data point corresponds to the $\ket{0}$ or $\ket{1}$ state, and choose the most probable option \cite{burrell2010high,shen2012correcting}. However, these methods are all tailored for a specific scenario and it is difficult to integrate other sources of information about the state, such as counts from extra PMT channels when imaging the ions onto alternating detectors, or photon arrival times. The latter contributes information about the state because bright-to-dark or dark-to-bright pumping events have characteristic photon arrival time distributions, i.e. photons arriving predominantly early or late in the detection window, which can be included in the discrimination procedure. To incorporate all  data sources in a single framework and reduce the effect of crosstalk we take advantage of advances in the field of machine learning, and use an artificial neural network to perform the discrimination task. Before proceeding to the main results, we briefly introduce the neural network framework that is used in this work.
\begin{figure}[t]
	\includegraphics[width=\linewidth]{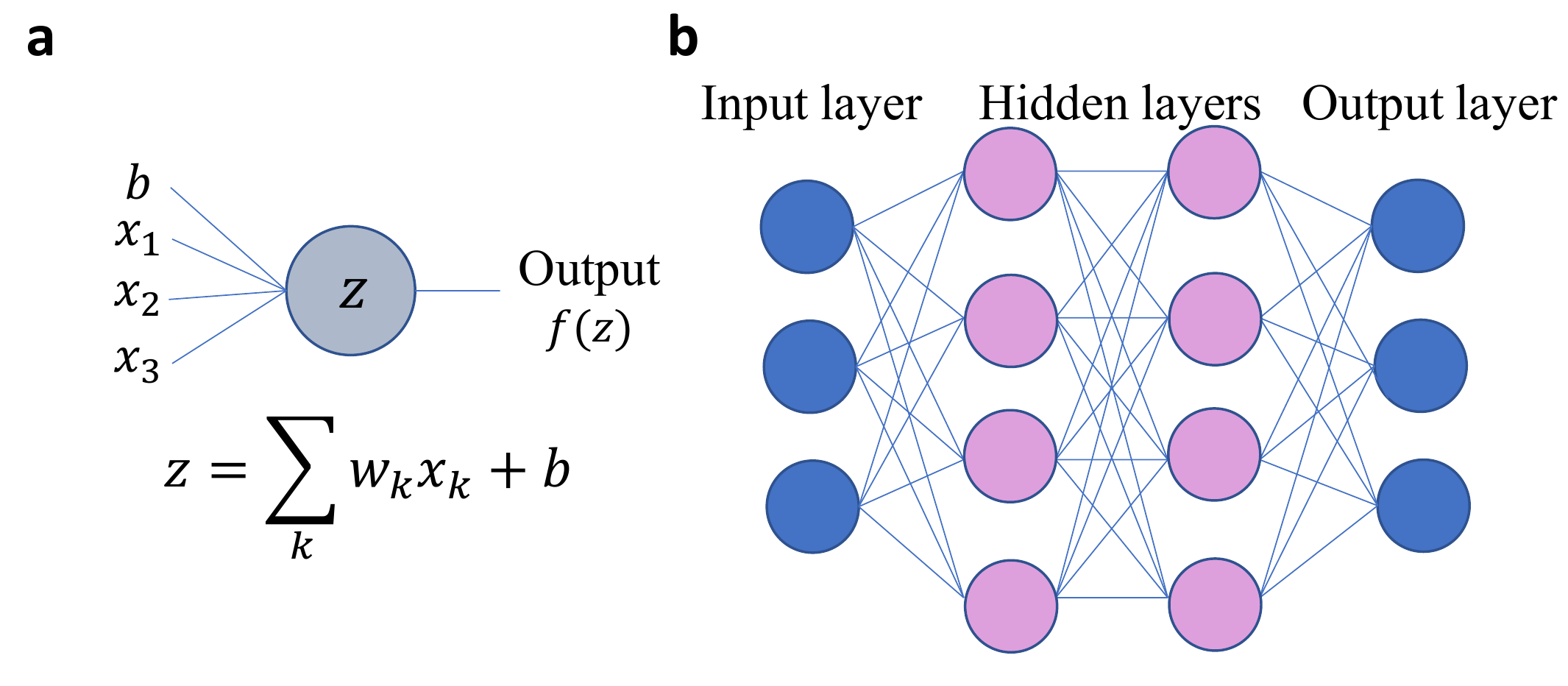}
	\caption{An artificial neuron and a neural network.  (a) A single neuron takes inputs $x_k$ and outputs $f(\sum_k w_k x_k +b)$, where $f$ is called the activation function, and $w_k$'s and $b$ are weights and the bias of the neuron. (b) A neural network is composed of artificial neurons stacked in layers and connected to each other. }
	\label{fig:neuralnet}
\end{figure}

\begin{figure}[t]
	\includegraphics[width=\linewidth]{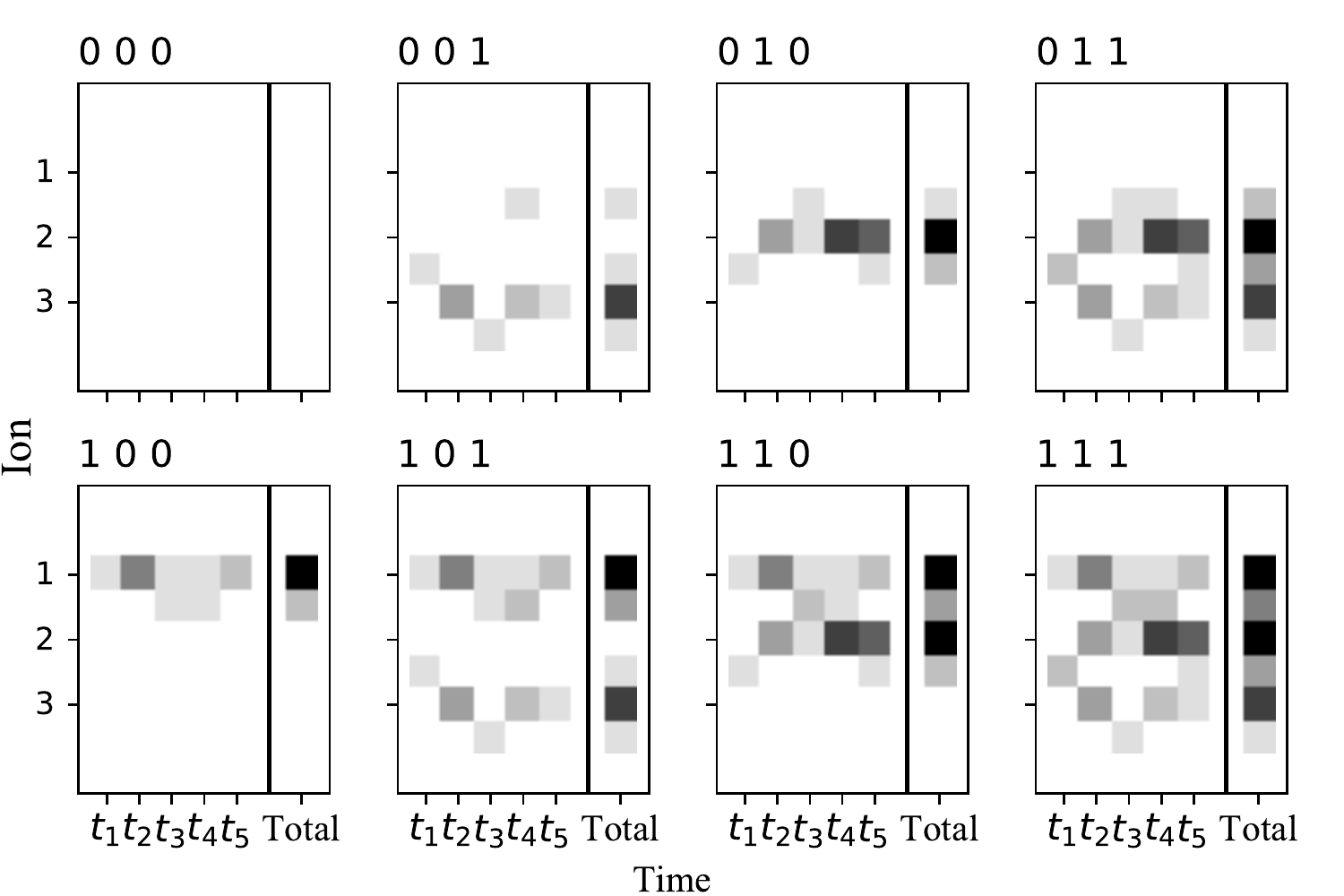}
	\caption{Results of an instance of the experiment for different initial states. Photon counts for all eight computation basis states over three qubits (000,001,\dots,111) are binned into 30 $\mu$s time-bins in a total of 150 $\mu$s  collection. The last column shows the total counts and darker color indicates more photons. The effect of crosstalk is visible in the intermediate channels (unlabeled rows). }
	\label{fig:PMT}
\end{figure}

With $N$ qubits, the measurement consists of photon counts and their arrival times on $M\geq N$ PMT channels. These photon counts can be binned into $T$ time-bins to give $M\times T$ numbers that completely describe the measurements. Our goal is to classify these measurement results into $2^N$ states, in an $N$-qubit basis. Therefore, we consider a supervised learning scenario, where a set of measurement results and their corresponding state labels is used to train the machine and predict the correct state corresponding to a given input.  We use a feed-forward neural network as depicted in Fig.~\ref{fig:neuralnet} \cite{lecun2015deep}. The network is built from a collection of neurons arranged in layers (columns in Fig.~\ref{fig:neuralnet}). A single neuron takes the inputs $x_k$, and outputs $f(\sum_k w_k x_k + b)$, where $f$ is called the activation function, and $w_k$ and $b_k$ are weights and biases respectively. One example of activation functions is the rectifier $f(z) = \max(0,z)$. %We can use a vectorized notation to denote the input to layer $i$ with $\mathbf{x}^{[i]}$. Therefore the weights and biases corresponding to the layer $i$ can be represented by a matrix $\mathbf{W}^{[i]}$ and a vector $\mathbf{b}^{[i]}$, respectively. In this notation, we have $\mathbf{x}^{[i+1]}=f(\mathbf{z}^{[i]} )$, where $\mathbf{z}^{[i]} =\mathbf{W}^{[i]}\mathbf{x}^{[i]} + \mathbf{b}^{[i]}$, and the function $f$ is applied element-wise.  
The first layer is called the input layer, where the neurons output the input data. Here, we have $M\times T$ neurons representing integrated photon counts from each ion in a time-bin (pixel values in Fig.~\ref{fig:PMT}). The last layer of the network is called the output layer. For classifying  data into exclusive classes, it is common to use a softmax activation at the output layer, i.e. $f_j(z_1,z_2,\dots z_k,\dots)=\frac{e^{z_j}}{\sum_k e^{z_k}}$. This choice of activation normalizes the output and can be interpreted as the probability of each input data belonging to each class $j$, and the output with the highest probability is chosen as the label $\tilde{y}$. These classes correspond to the $2^N$ different quantum states in our system (image labels in Fig~\ref{fig:PMT}). All the layers between the input and output are called hidden layers, and we use the rectifier function for them. We use a neural network with two hidden layers. The number of neurons in these layers varies from $8$ for the simplest case to $40$ for the network with the most features. The training task consists of finding weights and biases that optimize a cost function. We minimize the cross-entropy with the ADADELTA optimizer \cite{adadelta12}.
%\begin{equation}
%C = -\frac{1}{N_s} \sum_m [y_m \ln(\tilde{y}_m)  + (1-y_m) \ln(1-\tilde{y}_m)], 
%\end{equation}
%where the sum is carried over all the $N_s$ training samples, and $y_m$  and $\tilde{y}_m$ are the true label and the neural network prediction for the sample $m$, respectively.  

We now discuss the results in detail. We begin by moving a single trapped ion to the positions that would be occupied by ions in the multi-ion chain that we wish to investigate. This method allows us to  to recreate the experimental setup with $N$ qubits.  We typically image ions onto alternating PMT channels to reduce the crosstalk, which leaves the intermediate channels unused. We also take data imaging them onto neighboring PMT channels in order to explore how detection errors would change for a chain of ions with smaller inter-ion distance. Then, we either initialize the ion in $\ket{0}$ to take data on dark states or we use a high-fidelity microwave pi-pulse to create $\ket{1}$ for bright-state data. Finally, we detect the qubit state by counting how many photons are detected on the ion's corresponding PMT channel as well as neighboring channels. In addition, the photon arrival time is recorded with sub-\text{$\mu$s} resolution. By loading only a single ion, we can create the full statistics for all the $2^N$ computational basis states by superimposing these individual distributions. This procedure separates SPAM errors from other systematic errors present in the system such as addressing crosstalk errors. The average detection fidelity, which includes a small error from state preparation, is given by
\begin{equation}
\bar{\mathcal{F}} = 1/\mathcal{N} \sum_{i} p(\text{measured } i| \text{prepared }i)  ,
\end{equation}
where the sum is carried over all the computational basis states.

We compare six different methods and show that machine learning approaches outperform the two commonly used strategies in state discrimination. In this comparison, we train the neural network on 80\% of the data, and report the fidelities on the remaining 20\% test set.  Below we describe these six strategies: 
\begin{enumerate}[label=(\roman*)]
	\item Fixed threshold (\emph{FT}): A threshold for photon counts is chosen to maximize the discrimination between bright and  dark probability distributions. The same threshold is used for all the ions. In experiments with more than one qubit, this threshold is higher than the single qubit case because of crosstalk. Additional background noise from superimposing the statistic of individual qubits do not significantly contribute to errors.
	\item Adaptive threshold (\emph{AT}): The threshold for each ion depends on the state of its neighbors. First, the state is determined by a fixed threshold, and then the inference process is iterated based on the state of neighbors and the corresponding thresholds. 
	\item Neural network (\emph{NN}): First the photon counts from the ion PMT channels and their corresponding $2^N$ states (labels) are fed into a neural network. After the training, the neural network can predict the state of a given array of photon counts.
	\item Neural network with intermediate channels (\emph{NN+}): Similar to \emph{NN}, but the input also contains the intermediate PMT channel's data. 
	\item Neural network with time-stamped data (\emph{TNN}): The photon counts from the ion PMT channels are collected into time-bins to form a 2D image, where one axis is the time, and the other represents the location of the ions. The color intensity then represents the number of photons observed in that time-bin (see Fig.~\ref{fig:PMT}). These images with their corresponding labels are used to train the neural network.
	\item Neural network with time-stamped data and intermediate channels (\emph{TNN+}): The time-binned photon counts of the ion PMT channels and the intermediate channels are used to form an image, which the neural network learns to classify. This is the most comprehensive information available about the experiment.
\end{enumerate} 

We note that due to a large overlap between the photon count distributions of the intermediate channels with different bright neighbors, it is not possible to utilize the intermediate channel data with a simple threshold method. The same is true for the time-binned data, where the overlap of the bright and dark distributions is significant. This is because the distributions are Poissonian and have close mean values. However, the neural network can easily incorporate all the features and extract the available information.

\begin{figure}[t]
	\includegraphics[width=\linewidth]{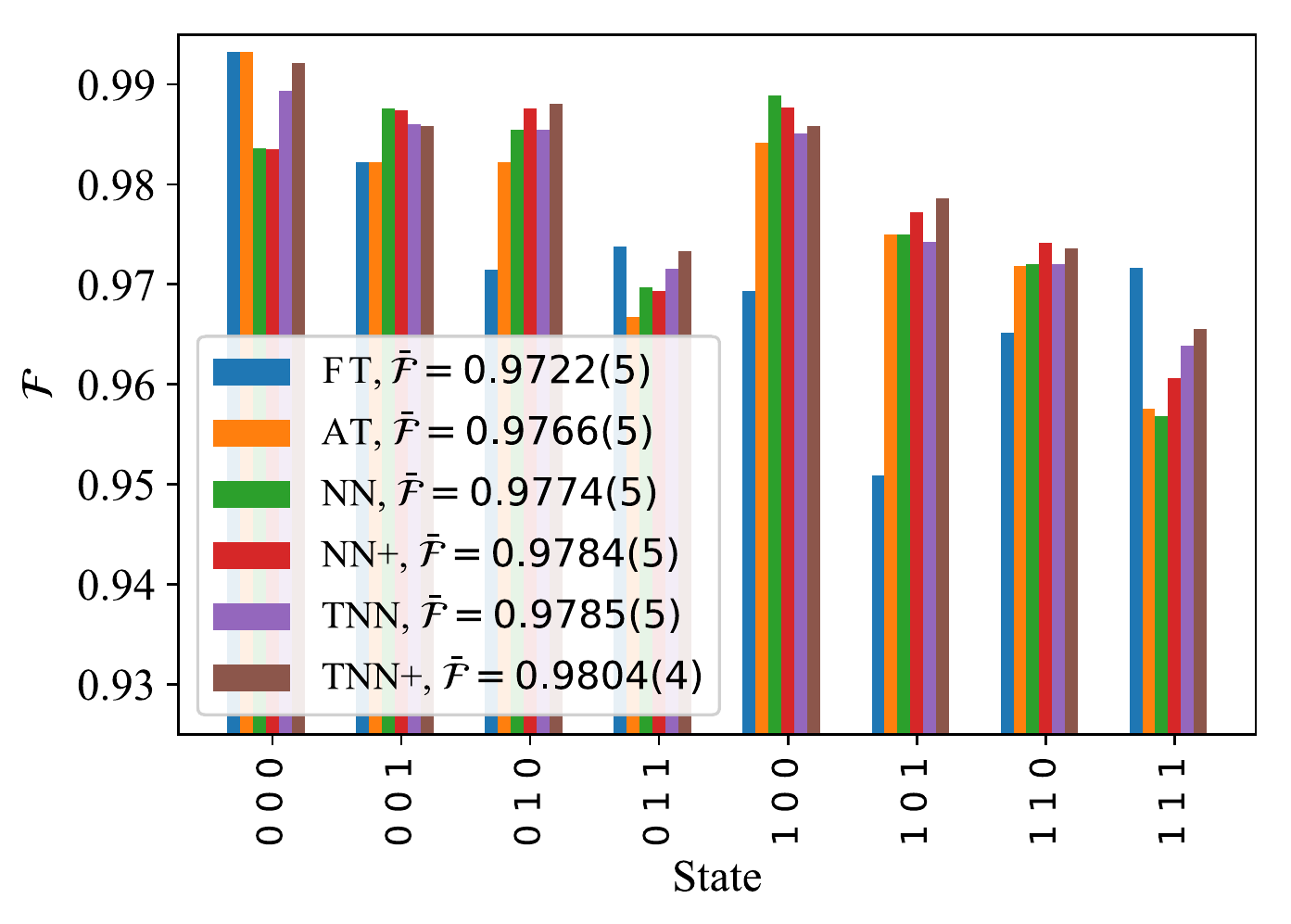}
	\caption{Comparison of different methods for state detection defined in the main body. We can see that the neural network (NN) methods outperform the conventional thresholding (\emph{FT}, \emph{AT}) methods. In addition, the performance is enhanced gradually as we provide the neural network with more features, e.g. intermediate channels and time stamps. The errors given in parentheses are statistical. }
	\label{fig:fidelitycomp}
\end{figure}

In the first experiment, we consider a three qubit measurement scenario where the data from intermediate channels is available. We collect $80000$ samples for each label and apply the six strategies and observe that the neural network outperforms the other methods. In  Fig.~\ref{fig:fidelitycomp}, it can be seen that with the same amount of information \emph{NN} outperforms \emph{T} and \emph{AT}, and when additional information is provided \emph{TNN+} can improve the errors over \emph{T} and \emph{AT} by 30\% and 17\% respectively. It can also be seen that the neural network reduces the false positives in detecting 000 and 111 states and improves the crosstalk errors in the other states. Note the architecture of the neural network is kept the same and only the number of neurons are increased to represent the more complicated features, therefore providing a flexible tool for inferring properties of the system from experimental data. 

In the second experiment, the ions are moved closer to each other to represent experiments where there are many ions in the trap, and neighboring PMT channels are associated with different ions. In this case, the data from intermediate channels is no longer available, and the crosstalk errors are increased. We consider a five qubit measurement scenario with $50000$ samples for each label and compare \emph{T}, \emph{AT},\emph{NN} and \emph{TNN} methods.
\begin{figure*}[t]
	\includegraphics[width=\linewidth]{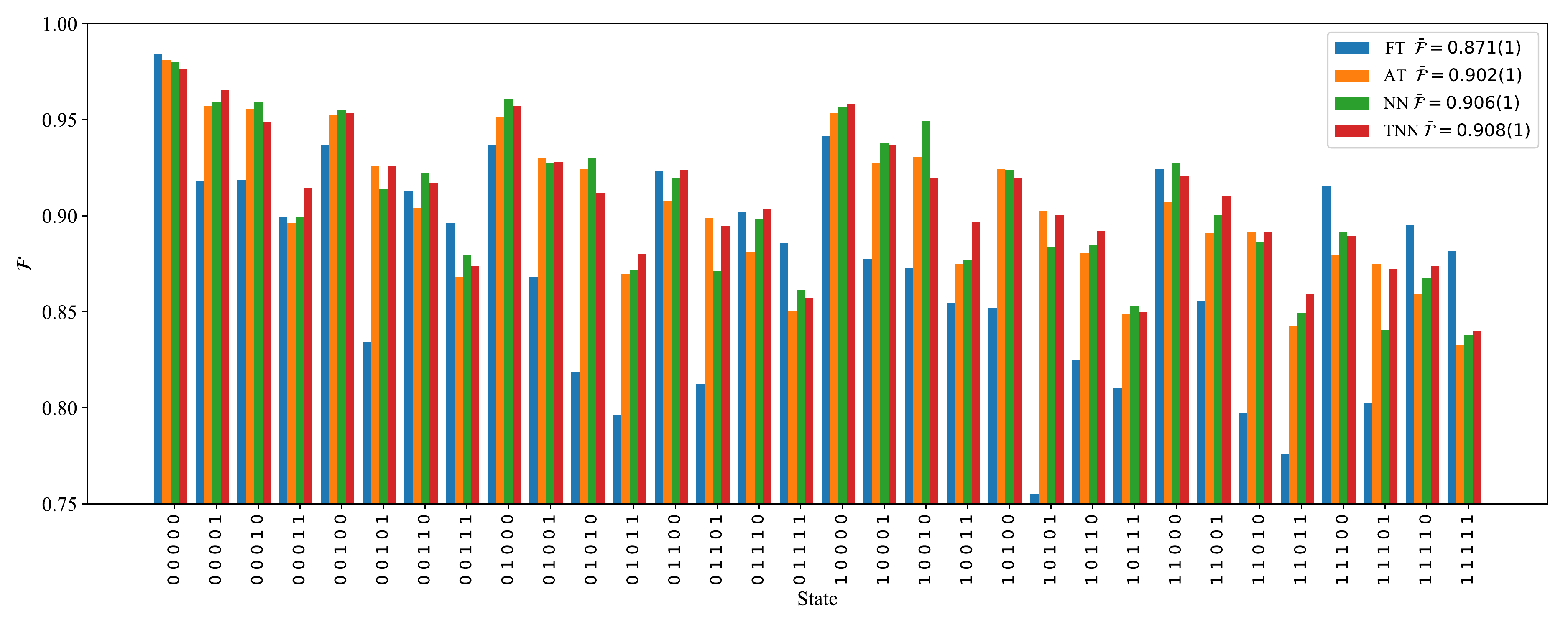}
	\caption{Comparing neural networks with threshold methods for five-qubit state detection. In this case the intermediate channel data is not available, but neural networks can still perform better than threshold methods. The errors given in parentheses are statistical.}
	\label{fig:fidelityclose}
\end{figure*}

As shown in Fig.~\ref{fig:fidelityclose} we can see that the same behavior observed in the first experiment persists, and neural networks beat threshold methods, and incorporating time stamped data further improves the fidelities. Specifically, we observe 29\% and 6\% improvement by \emph{TNN} over \emph{T} and \emph{AT}, respectively. 

\begin{figure}[t]
	\includegraphics[width=\linewidth]{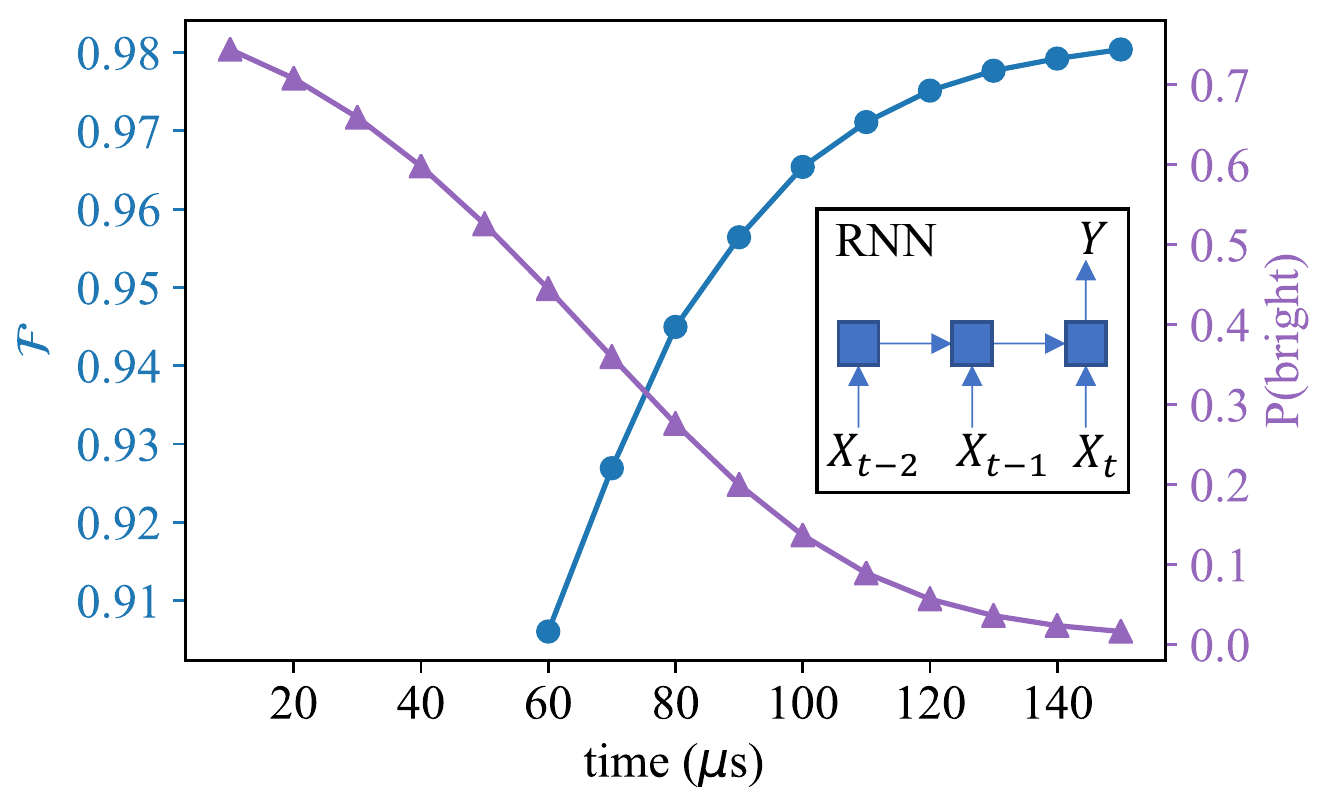}
	\caption{Recurrent neural network approach. The inset shows a schematic representation of the network, where the carried internal state and the output is affected by additional sequential inputs (time-binned photon counts). Left ordinate, blue circles: Performance of a recurrent network for different detection times. The fidelity increases with longer detection times. Right ordinate, purple triangles: The  average probability of the ion being bright decreases with the arrival time of the first photon. }
	\label{fig:figRNN}
\end{figure}

In addition,  we employ a recurrent neural network (RNN) as an alternative approach. These networks are tailored towards studying sequences of data (time-bins in our case), where the output in each step depends on the history through the internal state and an external input (see Fig.~\ref{fig:figRNN} inset). This feedback and memory effect is useful in capturing correlations in the sequence. While we observe the same fidelity as \emph{TNN+}, this method is advantageous for experiments with variable detection time, since it can handle data sequences with different lengths. We illustrate this capability by training a long short-term memory (LSTM) network, which is a type of RNN \cite{gers2002learning}, with the full sequence of measurement data, using finer time-bins of 10$\mu$s. Then, we evaluate the performance of the network by varying the length of the test sequence, and observe that indeed the performance increases with the measurement time (see circles in Fig.~\ref{fig:figRNN}). In addition, we interrogate the network with artificial data to map out its internal mechanism. Specifically, we construct sequences with a single photon count, the arrival time of which is scanned. The output indicates the significance of the photon arrival time in deciding the state of the qubits. We observe that the network learns that photons with late arrival times are more likely to come from ions prepared in the dark state, which is consistent with our physical understanding of error mechanism by off-resonant excitation (see traingles in Fig.~\ref{fig:figRNN}). 

We have shown that a simple neural network classifier can improve the detection fidelities over tradition thresholding methods. The neural network classifier does not require assumptions about the system and can incorporate different data sources in one framework in a straightforward way. As the ion-trap systems are very clean and the measurements are well-described theoretically, we do not expect  neural networks to beat complicated models that take into account possible errors and evaluate the likelihood of a state corresponding to measured values.  Similarly, we were not able to observe significant improvements over feed-forward networks by using RNNs or convolutional neural networks. This is because the patterns and correlations in the data are simple and hence well-captured by feed-forward networks. However, we expect our methods to be especially useful in other systems such as superconducting qubits where the measurement processes are more complicated and the data has intricate features.

In addition, we have considered exclusive labels in our classifier, which implies the size of the network scales exponentially with the number of qubits. However, we have observed that multi-label classifiers can achieve a performance close to our method, while maintaining a linear scaling with the number of qubits. Moreover, while it is not necessary in the current setup, by taking advantage of the locality of the crosstalk one can bootstrap smaller networks over a few qubits. By taking majority vote over the outcome of the smaller classifiers, the most probable state corresponding to the measurement results over many qubits is decided. 

In conclusion, we expect that techniques such as the one presented can simplify and improve the future experiments and serve as a straightforward alternative for optimizing the readout of quantum computers as they are scaled up to many qubits.\\

\section*{Acknowledgments}
AS thanks Evert van Nieuwenburg for helpful discussions. This work was supported by the IARPA LogiQ program, the ARO Atomic Physics program, the ARO and AFOSR MURI programs, Northrup Grumman, and the NSF Physics Frontier Center at JQI.

\bibliography{mybib}{}
\bibliographystyle{apsrev4-1}
\end{document}